\documentclass[10pt,showpacs,aps,prl,reprint,superscriptaddress]{revtex4-1}
\usepackage{graphicx}
\usepackage{multirow}
\usepackage{epstopdf}
\usepackage{amssymb}
\usepackage{dcolumn}
\usepackage{color}
\usepackage{amsmath}

\begin{document}

\title{Origin of the Counterintuitive Dynamic Charge in the Transition Metal Dichalcogenides}

\author{Nicholas A. Pike}
\email[]{Nicholas.pike@ulg.ac.be}
\affiliation{nanomat/Q-Mat/CESAM, Universit\'{e} de Li\`{e}ge \& European Theoretical Spectroscopy Facility, B-4000 Li\`{e}ge, Belgium}
 \author{Benoit Van Troeye}  
 \affiliation{Universit\'{e} Catholique de Louvain, Institute of Condensed Matter and Nanosciences 
(IMCN) \& European Theoretical Spectroscopy Facility, B-1348 Louvain-la-Neuve, Belgium}
 \author{Antoine Dewandre}
\affiliation{nanomat/Q-Mat/CESAM, Universit\'{e} de Li\`{e}ge \& European Theoretical Spectroscopy Facility, B-4000 Li\`{e}ge, Belgium}
 \author{Xavier Gonze}
 \affiliation{Universit\'{e} Catholique de Louvain, Institute of Condensed Matter and Nanosciences 
(IMCN) \& European Theoretical Spectroscopy Facility, B-1348 Louvain-la-Neuve, Belgium}
 \author{Matthieu J. Verstraete}
\affiliation{nanomat/Q-Mat/CESAM, Universit\'{e} de Li\`{e}ge \& European Theoretical Spectroscopy Facility, B-4000 Li\`{e}ge, Belgium}

\date{\today}

\begin{abstract}

We investigate the chemical bonding characteristics of the transition metal dichalcogenides based on their static and dynamical atomic charges within Density Functional Theory. The dynamical charges of the trigonal transition metal dichalcogenides are anomalously large, while in their hexagonal counterparts, their sign is even counterintuitive i.e. the transition metal takes the negative charge. This phenomenon cannot be understood simply in terms of a change in the static atomic charge as it results from a local change of polarization. We present our theoretical understanding of these phenomena based on the perturbative response of the system to a static electric field and by investigating the hybridization of the molecular orbitals near the Fermi level.  Furthermore, we establish a link between the sign of the Born effective charge and the $\pi$-backbonding in organic chemistry and propose an experimental procedure to verify the calculated sign of the dynamical charge in the transition metal dichalcogenides.
\end{abstract}

\pacs{31.15.ae- Electronic Structure and Bonding Characteristics, 63.20.dk - First-Principles Theory  }

\maketitle

The expanding interest in the Transition Metal Dichalcogenides (TMDs) stems from their wide variety of applications, ranging from batteries to electronic devices~\cite{Pumera2014,Jariwala2014,Tedstone2016}. Indeed, by adjusting either their chemical composition or the number of layers, one can tune their electronic, vibrational, and magnetic properties in a remarkable manner that cannot be imitated in other two-dimensional and layered materials. In particular, the TMDs offer high carrier mobilities~\cite{Gupta2015,Bhimanapati2015,Wang2012} and a strain-dependent indirect to direct band gap transition~\cite{Espejo2013}, that are crucial for future electronic and electro-optic applications~\cite{Radisavljevic2011,Kumar2013}. Still, while the electronic properties of these materials are now relatively well-known~\cite{Kuc2015}, the character of their chemical bonds is, interestingly, quite diverse, and, to the best of our knowledge, not yet fully understood.  For example, while ZrS$_2$ is reported as extremely ionic~\cite{White1972,Vaterlaus1985}, MoS$_2$ and WS$_2$ are reported to possess both ionic and covalent characteristics~\cite{Lucovsky1973,Li1996}. Additionally, TiS$_2$ was recently reported as metallic and semiconducting, both experimentally and theoretically~\cite{Fang1997,Liu2012,Sharma1999}. 

In this Letter, we aim to provide a deeper understanding of the bonding characteristics and charge transfer in the TMDs thanks to Density Functional Theory (DFT)~\cite{Martin2004}. One common way to estimate the charge distribution within this theory is to partition the electronic density of the system into constituent atoms~\cite{Bader1985,Hirshfeld1977}.   While conceptually simple, this notion of ``static'' charge is, unfortunately, ambiguous~\cite{Ghosez1998} and the corresponding charges cannot be measured experimentally. Contrarily, the Born effective charge (BEC)~\cite{Gonze1997b}, arising from the change of dipole moment due to an atomic perturbation, is a physical observable as it corresponds to the dynamical charge response to a perturbation. It governs, for example, the splitting between the transverse optical (TO) and longitudinal optical (LO) vibrational modes~\cite{Gonze1997b}. It has been argued~\cite{Meister1994} that all the various operational charge definitions (including static and dynamic charges) share a single principal component. However, in various materials, e.g. ferroelectric perovskites, the Born effective charge is observed to be anomalously large with respect to its static nominal value, albeit {\it without} a change in sign~\cite{Ghosez1998}.

In this paper, we highlight the critical differences between the dynamical BEC and static Bader charge in the cases of the hexagonal TMDs (h-TMDs). Indeed, our BEC calculations indicate that {\it the transition-metal atom takes the negative charge, while the nominal and computed static charges lead to the opposite conclusion}. Our sign and value agree with recent DFPT calculations in Refs.~\cite{Sohier2016, Danovich2017}, but disagree with the claimed sign (no values given) in Ref.~\citenum{Ataca2012}. The h-TMD contrast strongly with the trigonal TMDs (t-TMDs),  where we find that the signs of the Bader charge and of the BEC agree. In what follows, we discuss these opposite behaviors,  and more importantly, explain the origin of the counterintuitive sign of the BECs in the h-TMDs by investigating the localization of the hybridized molecular orbitals near the Fermi level.  Finally, we propose an experimental method to verify our theoretical observations.

Our calculations are performed using the {\it Abinit} software package~\cite{Gonze2005,Gonze2009,Gonze2016} with the GGA-PBE exchange-correlation functional~\cite{Martin2004,Perdew1996,Fuchs1999,Torrent2008,Marques2012}, corrected by Grimme's DFT-D3 functional for the dispersion corrections due to long range electron-electron correlations~\cite{Grimme2010,Troeye2016}.  With the inclusion of this van der Waals functional, we are able to reproduce the in-plane and out-of-plane experimental lattice parameters within 0.7$\%$~\cite{Dickinson1923,Brixner1962,Berkdemir2013,Traving1997,Wan2010,Chen2015}. Details on the norm-conserving pseudopotentials~\cite{blochl,JTH}, the convergence parameters (plane-wave expansion cutoff energy and Brillouin-zone sampling)  and the structural parameters is found in the Supplemental Material~\cite{supmat}.
\begin{table}[!t]
\centering
\begin{tabular}{c | c c | c c | c c }
\toprule
 \multirow{3}{*}{} &  \multicolumn{4}{c|}{Born effective charge [e]} & \multicolumn{1}{c}{Bader [e]} &\multicolumn{1}{c}{BPDC [e]}  \\ \cline{2-5} \cline{6-7}
 &  \multicolumn{2}{c|}{This work} & \multicolumn{2}{c|}{Exp.~\citenum{Sun2009,Wieting1980,Luttrell2006,Uchida1978,Vaterlaus1985}} &  &   \\ 
 & $Z^*_{\text{Mo},xx}$ & $Z^*_{\text{Mo},zz}$ & $|Z^*_{\text{Mo},xx}|$ & $|Z^*_{\text{Mo},zz}|$ &$Z^{B}_{\text{Mo,z}}$ &  $Z^{B,*}_{\text{Mo,z}}$ \\ \hline
 
MoS$_2$   & -1.090  &   -0.628  &  1.1 & 0.4 &  1.155 & 0.635  \\
MoSe$_2$  & -1.904 & -0.952 &2.1& 0.5& 0.910 & 0.652 \\
MoTe$_2$   & -3.280 & -1.562 & 3.4& & 0.575 & 0.752 \\ \hline
WS$_2$    & -0.505 & -0.426 & 0.4&0.2 &1.400 &  \\
WSe$_2$   & -1.242 &-0.776 & 1.7 &0.5 &1.081 &\\
\hline \hline
 & \multicolumn{6}{c}{ }\\ \hline
 TiS$_2$ &  ~6.344 & ~1.208 &6.0&2.2& 1.764 & 1.330 \\
 TiSe$_2$ &  ~8.230& ~1.092 &9.2&2.1& 1.599 \\          
                                                    \botrule                                     
\end{tabular}
\caption{\label{tab1} Computed BEC, static Bader charge, and BPDC for the transition metal atom in the h-TMDs and t-TMDs. The BPDC is defined in the main text. The BECs and Bader charges for the chalcogen atoms are exactly opposite and half the corresponding transition metal charge. The absolute value of the experimental BECs is also reported.}  
\end{table} 
 We investigate the Bader charge, $Z^B$, and BEC, $Z^*$, for the bulk MX$_2$ h-TMDs, where M=Mo, W, and X=S, Se, Te, as well as for TiS$_2$ and  TiSe$_2$, two semiconducting t-TMDs. We use Density Functional Perturbation Theory (DFPT)~\cite{Baroni2001,Gonze1997a,Gonze1997b} to calculate the BECs with the charge neutrality condition imposed.  All calculations of the static and dynamic charges use the relaxed geometries for the individual compounds.

In Table~\ref{tab1}, we report our calculated BECs and Bader charges for the previously-introduced TMDs, alongside experimental data extracted from infrared spectra~\cite{Sun2009,Wieting1980,Luttrell2006,Uchida1978,Vaterlaus1985}. These experiments only provide a measure of the magnitude of the BEC, and our calculated BECs must be compared accordingly. For both h-TMDs and t-TMDs, our computed BECs compare relatively well with the available experimental data. However, the BECs are anomalously large in t-TMDs, as they differ strongly from both their corresponding nominal and static charges~\cite{Ghosez1998}, while, in h-TMDs, we observe that the dynamical charges are counterintuitive, with the transition metal and chalcogen atoms taking the negative and positive charges, respectively, in disagreement with the corresponding nominal and static charges. These counterintuitive BECs for h-TMDs were also observed recently, even for the monolayer TMDs~\cite{Sohier2016,Danovich2017}, although the authors did not provide explanation for the calculated sign. To our knowledge this is the first clear case in which the sign of the static charge and BEC disagree, with the absolute difference being more than three electronic charges in the extreme case of MoTe$_2$, in contrast with the early belief~\cite{Meister1994} that all the various operational charge definitions share a single principal component.

In the following, we will first discuss the discrepancies observed between the static and dynamical charges in the h-TMDs, before explaining the physical origin of the counterintuitive sign of the BECs.

One has to remember that the static and dynamical charges cannot be directly compared as they represent different physical quantities; the static charge corresponds to a partition of the ground-state electronic density, while the dynamic charge corresponds directly to the dynamic response due to an atomic perturbation. Still, one can construct a dynamical charge based on the static charge by taking into account the change of Bader charge with an atomic displacement~\cite{Ghosez1998} computed by finite differences. This newly-constructed Bader Partitioned Dynamic Charge (BPDC), denoted $Z^{B,*}$ includes additional effects i.e. the charge (de)localization.

\begin{figure}[!t] 
    \centering
    \includegraphics[width=0.5\textwidth]{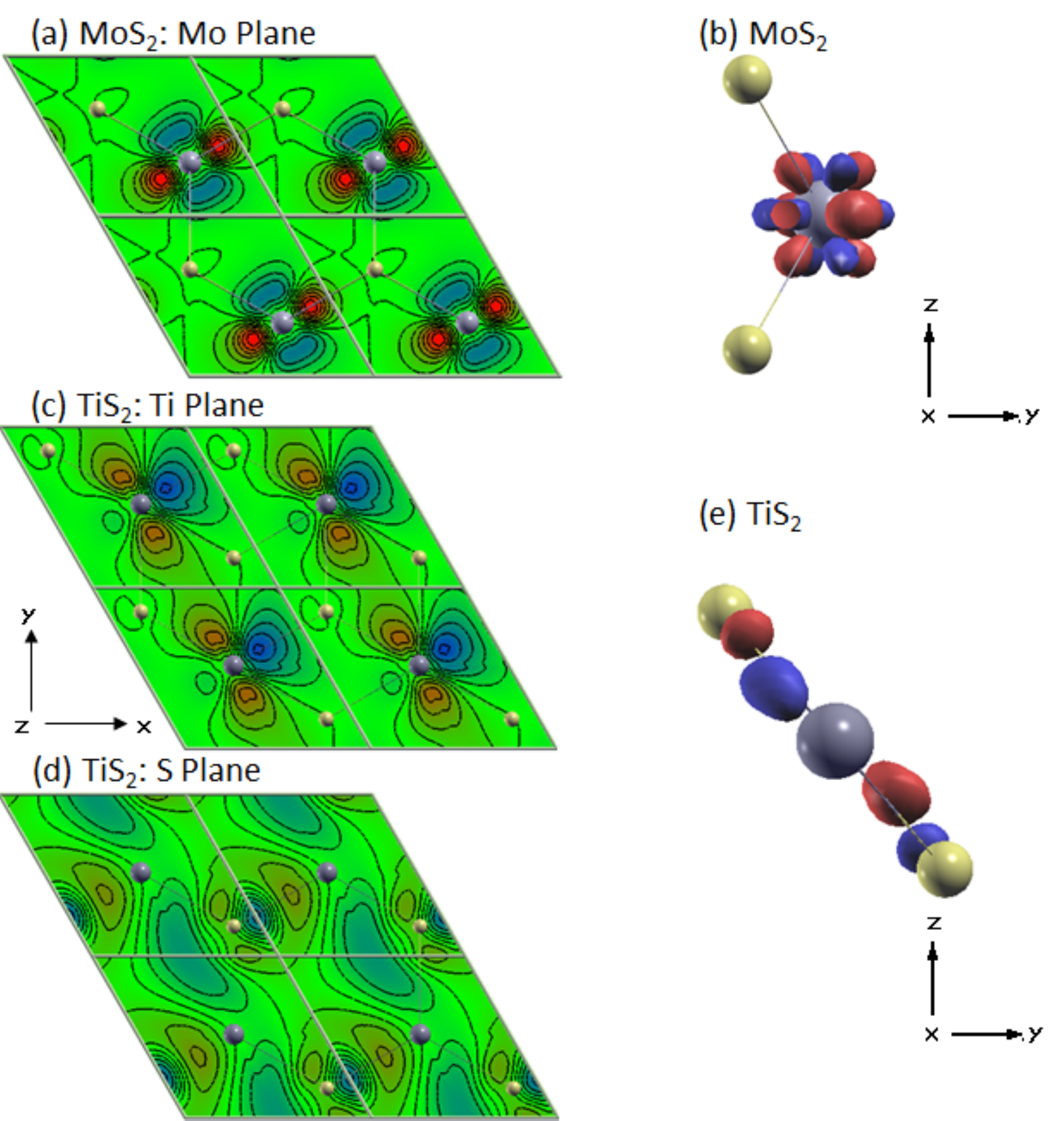} 
  \caption{ \label{change_den} (Color Online) Contour plots and isosurfaces of the change in the electronic density due to an electric field perturbation along the $x$ axis for MoS$_2$ and TiS$_2$.  For MoS$_2$ we show, (a) a contour plot of the in-plane change in the electronic density around a Mo atom and (b) the isosurface of the induced change in the electronic density for a single formula unit of MoS$_2$. The change in the electronic density around S is negligible (not presented). For TiS$_2$ we show the contour plots of the in-plane change in electronic density around (c) Ti and (d) S, as well as (e) the isosurface of the induced change in the electronic density for a single formula unit of TiS$_2$. Blue, red and green indicates positive, negative, negligible changes in the electronic density. The atomic colors indicate the transition metal atom, in gray, and the chalcogen atom, in tan.}
\end{figure}

In plane, the displacement of an atom can generate both a charge transfer and an electron current. For simplicity, we examine here atomic displacements in the out-of-plane direction, for which we assume the corresponding electron current to be zero due to the large distance between the layers. The corresponding charges are reported in Table~\ref{tab1}. While this dynamic correction to the Bader static charges is negative in most cases, in agreement with the sign of  $\Delta Z$, it is clearly too small to fully explain the sign of the BECs in h-TMDs. We explain the counterintuitive charges in terms of a local change of polarization around the atoms, that cannot be quantified by a partitioning approach~\cite{Ghosez1998}. This is confirmed by the analysis of the perturbed density with respect to an electric field perturbation, presented in Fig.~\ref{change_den} which is localized close to the Mo atoms. Changes within this region cannot be quantified by the Bader approach, in contrast to TiS$_2$ where most contributions come from outside the Ti Bader volume.

Consequently, a more direct analysis of the BECs is crucial to understand the discrepancies in the dynamical charges between the h-TMDs and the t-TMDs. In what follows, we focus on MoS$_2$ and TiS$_2$. The band-by-band decomposition~\cite{Ghosez2000} and localization tensor~\cite{Veithen2002} are unable to bring any simple or conclusive explanations on the difference of BECs between these two materials as shown in Table~S2 of the Supplemental Material~\cite{supmat}. Thus, it is necessary to analyze the different contributions to the BECs which are given explicitly, for example, in Ref.~\citenum{Gonze1997b}. Neglecting the separable part, the dynamic screening component, given in Eq.~(S3) of the Supplemental Material~\cite{supmat}, depends on, first, the change in the electronic wavefunction due to an electric field perturbation and, second, on the change of electronic potential due to an atomic displacement. While the change in potential does not vary qualitatively between MoS$_2$ and TiS$_2$ (see Fig.~S2 of the supplemental material~\cite{supmat}), their first-order density responses differ significantly, as illustrated in Fig.~\ref{change_den}. In MoS$_2$, this change of electronic density with an external electric field is localized around the Mo atom (Fig.~\ref{change_den}b) and takes a hybridized d-orbital shape, while, in TiS$_2$, it is delocalized along the Ti-S bond (Fig.~\ref{change_den}e). 

This localization/delocalization of the electronic density responses in MoS$_2$ and TiS$_2$, which results in the opposite character of the BEC between these materials, should arise from a different orbital hybridization and electronic configuration of h-TMDs and t-TMDs. With this in mind, and in order to understand the fundamental differences in the orbital hybridization in MoS$_2$ and TiS$_2$, we present a Molecular Orbital (MO) diagram~\cite{Fleischauer1989} based on the previous work of Stiefel {\it et al.}~\cite{Stiefel1966}. Similar to their work, we write down the molecular orbitals of MoS$_2$ and TiS$_2$ monolayers using the irreducible representation of the molecular orbitals for a single formula unit within these compounds.  Therefore, we use the point group symmetries $D_{3h}$ and $D_{3d}$, for MoS$_2$ and TiS$_2$, respectively. The orbital energy ordering was obtained by a direct comparison to the projected band analysis of MoS$_2$ and TiS$_2$ presented in the Supplementary Material~\cite{supmat}.

\begin{figure}[!t]
\includegraphics[width=0.49\textwidth]{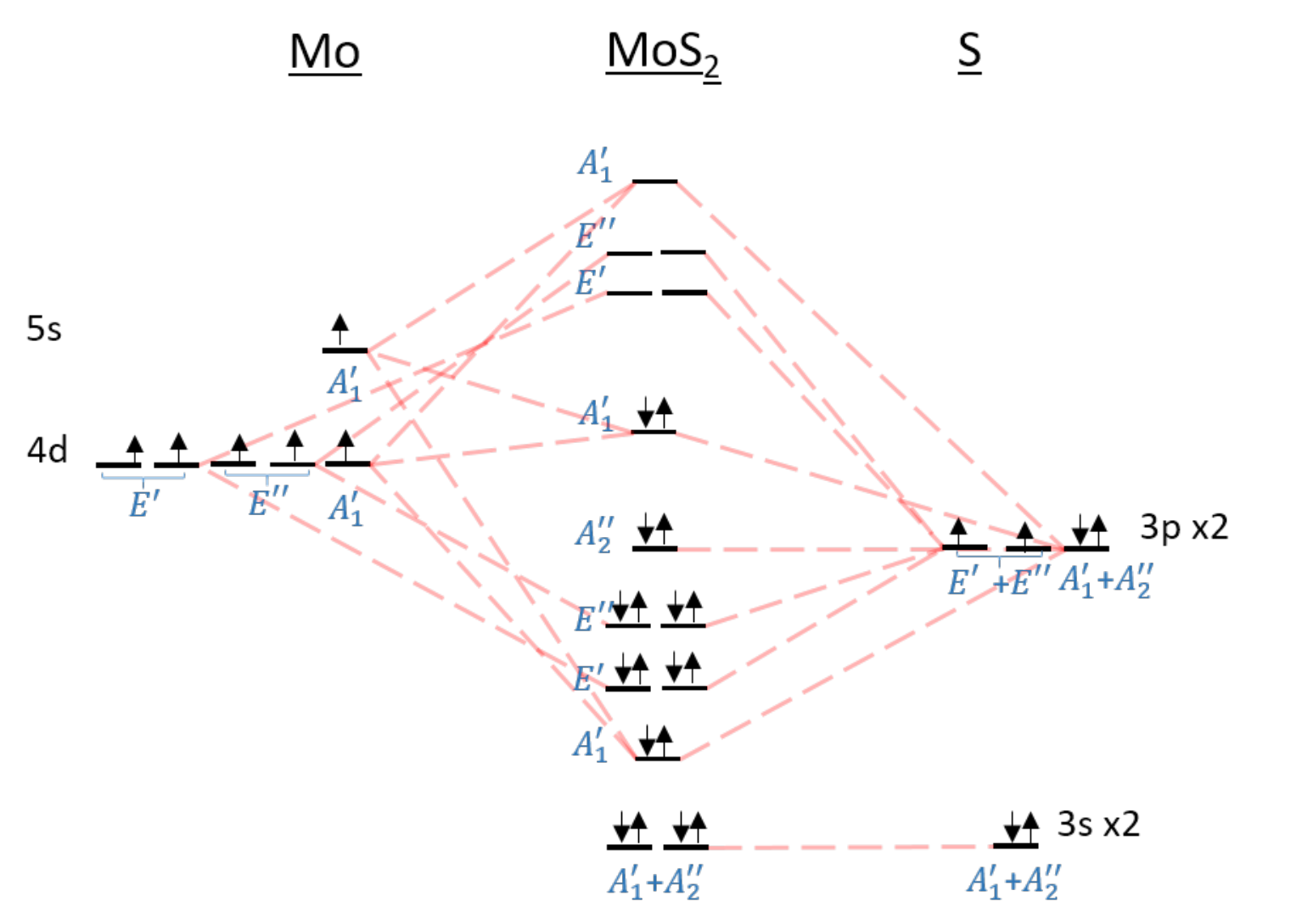}
\caption{\label{MO_diagram} (Color Online) Molecular orbital diagram of MoS$_2$. The symmetry notation, in blue, labels the symmetry type of the molecular orbital and the light red dashed lines are the symbolic links between the atomic and molecular orbitals.}
\end{figure}

The MO diagram of MoS$_2$ is presented in Fig.~\ref{MO_diagram}. It indicates that the lowest A$^\prime_1$, as well as the lowest E$^\prime$ and E$^{\prime\prime}$ molecular orbitals of MoS$_2$, are all bonding orbitals. According to the projected orbital analysis, this A$^\prime_1$ is mostly of S character, while the E$^\prime$ and E$^{\prime\prime}$ share both Mo and S orbital characteristics. The A$^{\prime\prime}_2$ orbital, arising from the interaction between p$_{z}$ orbitals of S, does not hybridize with the Mo atomic orbitals. The last occupied orbital -the A$^\prime_1$ orbital-, lies closest to the Fermi energy, and is an antibonding orbital arising from Mo orbitals, with a small amount of S component. The first unoccupied states correspond to the anti-bonding states E$^\prime$ and E$^{\prime\prime}$ that exhibit both Mo and S orbital characteristics.

\begin{figure}[t] 
    \centering
     \includegraphics[width=0.5\textwidth]{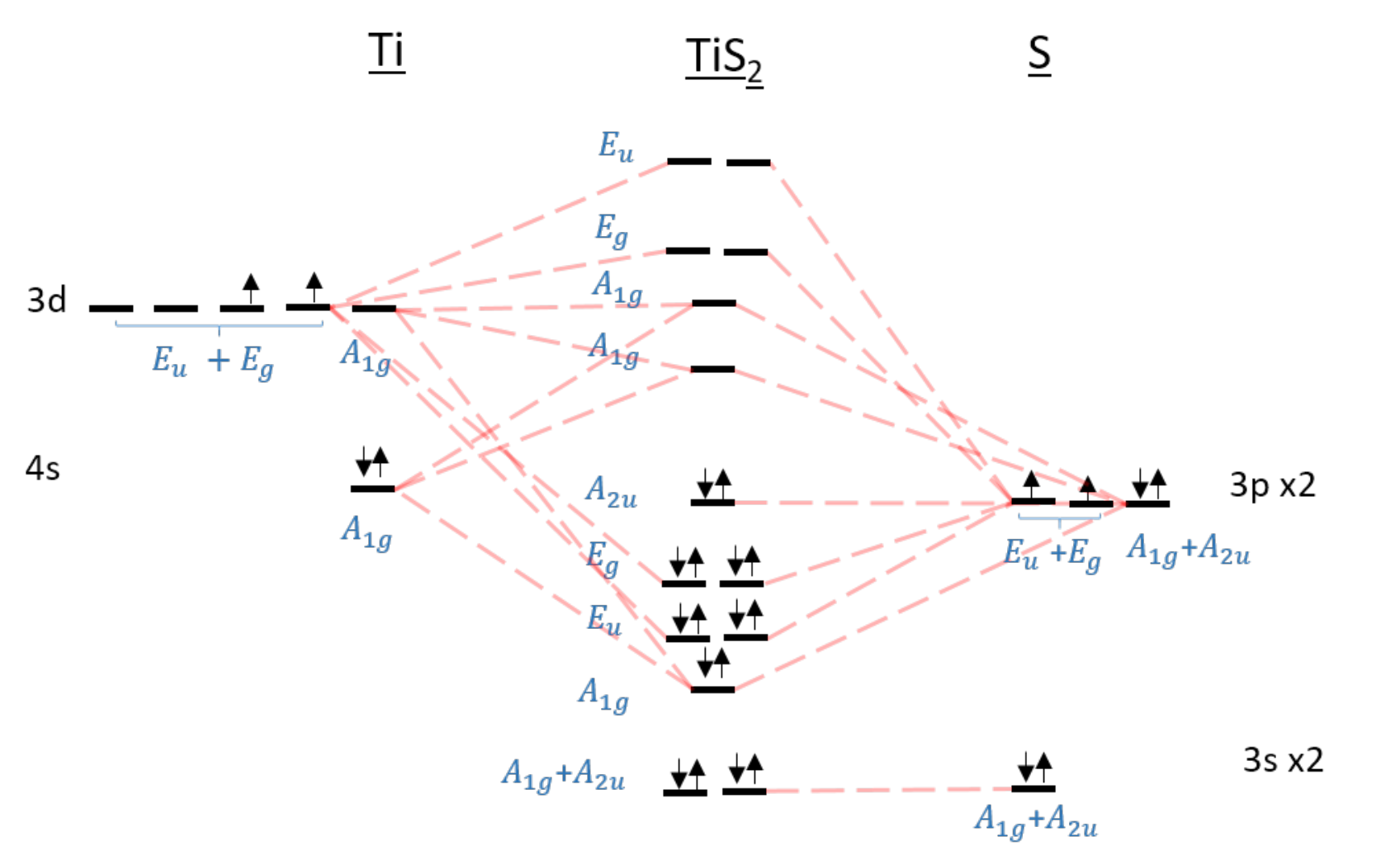} 
  \caption{ \label{MO_diagram_TI}(Color Online) Molecular orbital diagram of TiS$_2$. The symmetry notation, in blue, labels the symmetry type of the molecular orbital and the light red dashed lines are the symbolic links between the atomic and molecular orbitals.}
\end{figure}

The MO diagram of TiS$_2$ is depicted in Fig.~\ref{MO_diagram_TI} where we find that all the valence bands of TiS$_2$ contain a majority of atomic-like S states, in agreement with the projected band analysis in the supplemental material~\cite{supmat}, indicating charge transfer from Ti to S. Especially, note the inversion of the atomic Ti~4s and 3d states, compared to MoS$_2$, and the bonding character of the highest occupied MO.

For the case of MoS$_2$, we can make a parallel with the $\pi$-backbonding effect in organic chemistry, a process in which not only a $\sigma$ bond forms between a metal atom and a ligand, but also an additional $\pi$ bond that involves an antibonding state of the ligand~\cite{Miessler1999}. This second bond transfers charge back to a d-state of the metal atom, leading to a weakening of the ligand internal bond. In the case of MoS$_2$, the Mo atoms share their electrons with the S atoms, that transfer back their electrons to the Mo atoms (antibonding state of p$_z$ orbitals in destructive phase) in order to fill the 4d$_{z^2}$ orbital of Mo. The presence of two types of atomic orbitals from the transition metal, s and d, with different spatial extents, is critical in both the well-known $\pi$-backbonding phenomena and the present counterintuitive and anomalous BECs. 

With these MO diagrams in mind, and focusing on the out-of-plane direction, we are now able to understand the origin of the BECs sign in the h-TMDs. Compared to the monolayer, the HOMO level of bulk h-TMDs (now split due to the AB stacking of the TMDs) remains antibonding. This orbital is rather localized around Mo due to its antibonding character, in contrast to the other bonds in this compound, that are found to be mostly delocalized (with Mo and S characteristics). However, this localized bond, corresponding to a superposition of Mo 4d$_{z^2}$, Mo $5s$ and S $3p$ states is especially sensitive to atomic displacements.  This bond gives rise to the local change in polarization that was described previously in this Letter and is itself responsible for the counterintuitive sign of the BECs in the h-TMDs. On the contrary, the last occupied orbitals of TiS$_2$ are all bonding and delocalized, and thus do not lead to any local change of polarization. The explanation remains valid for monolayer h-TMDs, and for the in-plane components of the BECs as well (we find the same anomalous sign in all cases).

Experimentally it may be difficult to determine the sign of the BEC, as most relevant experimental quantities, in particular, the LO-TO splitting, depend on the magnitude of the BEC and not its sign. However, since the BEC is an observable quantity (e.g. the polarization when atoms are displaced), its sign should be measurable, e.g. in a system where the mirror plane symmetry is broken. A small movement of the transition metal ion would then lead to an asymmetric effect as a function of applied fields or strain. There exist other layered TMD materials similar to the h-TMDs, but with lower lattice symmetry due to stacking, in particular, TcS$_2$, ReS$_2$, and ReSe$_2$ which belong to a triclinic space group.  Their unit cells are more complex, with inequivalent metal sites and buckled chalcogens.  
In these cases, the Raman susceptibility tensor can be used to determine the sign of the BEC in these materials: the components of the Raman susceptibility tensor are \emph{linearly} dependent on the BEC~\cite{Veithen2005} and can be used to deduce the sign a BEC in an angle-resolved Raman measurement similar to Wolverson {\it et al.}~\cite{Wolverson2014} an example of which is given in the supplemental material~\cite{supmat}. Finally, it may also be possible to measure the inverted sign of the BEC using X-ray absorption spectroscopy in a strong electric field, which is atom-specific~\cite{Ney2016}. Note that, for h-TMD in a homogeneous field of any direction, half of the bonds will be stretched and the other half compressed, making it impossible to distinguish the BEC sign.

In conclusion, we have highlighted the counterintuitive sign of the Born Effective Charge in the hexagonal TMDs. This sign derives from an important local change of polarization around the transition metal atom, caused by an antibonding occupied orbital close to the Fermi level  involving the $d$ electrons of the transition metal and the $p$ electrons of the chalcogens. Interestingly, such chemistry is shared by a many compounds, but all of them do not show counterintuitive Born effective charges. A high-throughput screening is underway and may bring to light more specific requirements. We make a parallel with the $\pi$-backbonding effect in organic chemistry, and propose methods to confirm the sign of the computed BECs experimentally.

\section*{Acknowledgments}
We gratefully acknowledge discussions with G.~Petretto and Ph.~Ghosez. The authors acknowledge the Belgian Fonds National de la Recherche Scientifique FNRS under grant number PDR T.1077.15-1/7 (N.A.P and M.J.V) and for a FRIA Grant (B.V.T.). M.J.V and A.D. acknowledge support from ULg and from the Communaut\'{e} Fran\c{c}aise de Belgique (ARC AIMED 15/19-09). Computational resources have been provided by the Universit\'{e} Catholique de Louvain (CISM/UCL); the Consortium des Equipements de Calcul Intensif en F\'{e}d\'{e}ration Wallonie Bruxelles (CECI), funded by FRS-FNRS G.A. 2.5020.11; the Tier-1 supercomputer of the F\'{e}d\'{e}ration Wallonie-Bruxelles, funded by the Walloon Region under G.A. 1117545; and by PRACE-3IP DECI grants, on ARCHER and Salomon (ThermoSpin, ACEID, OPTOGEN, and INTERPHON 3IP G.A. RI-312763).


\end{document}